%
%
%
\def\today{\ifcase\month\or January\or February\or March\or April\or May\or
June\or July\or August\or September\or October\or November\or December\fi
\space\number\day, \number\year}
%
%
\newcount\notenumber

\def\note{\global\advance\notenumber by 1 \footnote{$^{\the\notenumber}$}}
%
%
\newif\ifsectionnumbering
\newcount\eqnumber
\def\cleareqnumber{\eqnumber=0}
\def\numbereq{\global\advance\eqnumber by 1
\ifsectionnumbering\eqno(\the\secnumber.\the\eqnumber)\else\eqno
(\the\eqnumber)\fi}
\def\eqalinno{{\global\advance\eqnumber by 1}
\ifsectionnumbering(\the\secnumber.\the\eqnumber)\else(\the\eqnumber)\fi}
\def\name#1{\ifsectionnumbering\xdef#1{\the\secnumber.\the\eqnumber}
\else\xdef#1{\the\eqnumber}\fi}
\def\nosectionnumbering{\sectionnumberingfalse}
\sectionnumberingtrue
%
%
\newcount\refnumber

\immediate\openout1=refs.tex
\immediate\write1{\noexpand\frenchspacing}
\immediate\write1{\parskip=0pt}
\def\ref#1#2{\global\advance\refnumber by 1%
[\the\refnumber]\xdef#1{\the\refnumber}%
\immediate\write1{\noexpand\item{[#1]}#2}}
\def\tie{\noexpand~}

%
%
\font\twelvebf=cmbx10 scaled \magstep1
\newcount\secnumber

\def\newsection#1.{\ifsectionnumbering\cleareqnumber\else\fi%
	\global\advance\secnumber by 1%
	\bigbreak\bigskip\par%
	\line{\twelvebf \the\secnumber. #1.\hfil}\nobreak\medskip\par\noindent}
%
%
%
\def \sqr#1#2{{\vcenter{\vbox{\hrule height.#2pt
	\hbox{\vrule width.#2pt height#1pt \kern#1pt
		\vrule width.#2pt}
		\hrule height.#2pt}}}}

%
%
%
\newdimen\fullhsize
\def\fiddle{\fullhsize=6.5truein \hsize=3.2truein}
\def\fullline{\hbox to\fullhsize}
\def\mkhdline{\vbox to 0pt{\vskip-22.5pt
	\fullline{\vbox to8.5pt{}\the\headline}\vss}\nointerlineskip}
\def\mkftline{\baselineskip=24pt\fullline{\the\footline}}
\let\lr=L \newbox\leftcolumn
\def\twocolumns{\fiddle
	\output={\if L\lr \global\setbox\leftcolumn=\columnbox
		\global\let\lr=R \else \doubleformat \global\let\lr=L\fi
		\ifnum\outputpenalty>-20000 \else\dosupereject\fi}}
\def\doubleformat{\shipout\vbox{\mkhdline
		\fullline{\box\leftcolumn\hfil\columnbox}
		\mkftline} \advancepageno}
\def\columnbox{\leftline{\pagebody}}
\nosectionnumbering
\magnification=1200
\def\pr#1 {Phys. Rev. {\bf D#1\tie }}
\def\pe#1 {Phys. Rev. {\bf #1\tie}}
\def\pre#1 {Phys. Rep. {\bf #1\tie}}
\def\pl#1 {Phys. Lett. {\bf #1B\tie }}
\def\prl#1 {Phys. Rev. Lett. {\bf #1\tie }}
\def\np#1 {Nucl. Phys. {\bf B#1\tie }}
\def\ap#1 {Ann. Phys. (NY) {\bf #1\tie }}
\def\cmp#1 {Commun. Math. Phys. {\bf #1\tie }}
\def\imp#1 {Int. Jour. Mod. Phys. {\bf A#1\tie }}
\def\mpl#1 {Mod. Phys. Lett. {\bf A#1\tie}}
\def\jhep#1 {JHEP {\bf #1\tie}}
\def\nuo#1 {Nuovo Cimento {\bf B#1\tie}}
\def\cla#1 {Class. Quant. Grav. {\bf #1\tie }}
\def\jetp#1 {JETP Lett. {\bf #1\tie}}
\def\tie{\noexpand~}

\parskip=15pt plus 4pt minus 3pt
\headline{\ifnum \pageno>1\it\hfil Linearized Analysis of
the Dvali-Gabadadze-Porrati model $\ldots$\else \hfil\fi}
\font\title=cmbx10 scaled\magstep1
\font\tit=cmti10 scaled\magstep1
\footline{\ifnum \pageno>1 \hfil \folio \hfil \else
\hfil\fi}
\raggedbottom


\overfullrule0pt


\rightline{\vbox{\hbox{RU01-16-B}\hbox{hep-th/0111127}}}
\vfill
\centerline{\title LINEARIZED ANALYSIS OF THE DVALI-GABADADZE-PORRATI}
\bigskip
\centerline{\title BRANE MODEL }
\vfill
{\centerline{\title Ioannis Giannakis
and Hai-cang Ren \footnote{$^{\dag}$}
{\rm e-mail: \vtop{\baselineskip12pt
\hbox{giannak@summit.rockefeller.edu, ren@summit.rockefeller.edu,}}}}
}
\medskip
\centerline{{\tit Physics Department, The Rockefeller
University}}
\centerline{\tit 1230 York Avenue, New York, NY
10021-6399}
\vfill
\centerline{\title Abstract}
\bigskip
{\narrower\narrower
We solve the Einstein equations in the Dvali-Gabadadze-Porrati
model with
a static, spherically symmetric matter distribution on
the {\it physical brane} and obtain an exact
expression for the gravitational field outside the source to
the first order in the gravitational coupling. Although, this expression
when confined on the {\it physical brane}
reproduces the correct $4-D$ Newtonian potential for distances
$r_s \ll r \ll {\lambda}$, where $\lambda$ is a characteristic
length scale of the model, it does not coincide
with the standard linearized form of the $4-D$ Schwarzschild metric.
The solution reproduces the $5-D$ Schwarzschild
metric in the linearized approximation for $r \gg {\lambda}$.
\par}
\vfill\vfill\break



Recently there have been several proposals which attempt to
achieve localization of gravity. The Randall-Sundrum model
\ref{\randall}{
L. Randall and R. Sundrum, \prl83 (1999) 4690.}
is based on the assumption that ordinary
matter and its gauge interactions are confined within
a four-dimensional hypersurface embedded in a five-dimensional
space which is {\it asymptotically anti-de Sitter}.
Since gravity is formulated by five-dimensional general relativity,
the gravitational field produced by a mass source on the brane
might deviate from the standard Schwarzschild metric.
Calculation of the form of the metric produced by a static,
spherical mass distribution on the brane, to first order
\ref{\giddings}{S. Giddings, E. Katz
and L. Randall, \jhep0003 (2000) 023.}
and to second order \ref{\ren}{I. Giannakis and H. C. Ren, \pr63 (2001)
024001.}, \ref{\kudoh}{H. Kudoh and T. Tanaka, \pr64 (2001) 084022.},
in the gravitational coupling revealed
no tangible disagreement with the classical tests of general
relativity. The Randall-Sundrum model was shown also to be consistent
with the gravitomagnetic effect \ref{\gma}{A. Nayeri
and A. Reynolds, {\it `` Gravitomagnetism in Brane Worlds ''},
hep-th/0107201.}. Various
aspects of black hole physics in the framework of the brane world
have been discussed in \ref{\black}{A. Chamblin, S. W. Hawking
and H. S. Reall, \pr61 (2000) 065007, T. Shiromizu and M. Shibata,
\pr62 (2000) 127502, N. Dadhich, R. Maartens, P. Papadopoulos
and V. Rezania, \pl487 (2000) 1,
A. Chamblin, H. Reall, H. Shinkai and T. Shiromizu,
\pr63 (2001) 064015,
R. Emparan,
G. Horowitz and R. Myers, \jhep0001 (2000) 007; \prl85 (2000) 499,
J. Garriga and M. Sasaki, \pr61 (2000) 065007,
I. Giannakis and H. c. Ren, \pr64 (2001) 065015,
I. Oda, {\it ``Reissner-Nordstrom Black Hole in Gravity Localized
Models ''}, hep-th/0008055,
M. Bruni, C. Germani and R. Maartens, {\it `` Gravitational
Collapse on the Brane''}, gr-qc/0108013, 
S. Vacaru and E. Gabunov,
{\it `` Anisotropic Black Holes in Einstein and Brane Gravity''},
hep-th/0108065, M. S. Modgil, S. Panda and G. Sengupta, {\it
`` Rotating Brane World Black Holes''}, hep-th0104122,
W. T. Kim, J. Oh, M. K. Oh and M. S. Yoon, {\it `` Brane World Black
Holes in Randall-Sundrum models ''}, hep-th/0006134,
P. Kanti and K. Tamvakis, {\it `` Quest for Localized 4-D Black
Holes in Brane Worlds''}, hep-th/0110298, T. Wiseman,
{\it `` Relativistic Stars in
Randall-Sundrum Gravity''}, hep-th/0111057.}.

An alternative gravity localized model,
the Dvali-Gabadadze-Porrati model \ref{\pora}
{G. Dvali, G. Gabadadze and M. Porrati \pl485
(2000) 208.}
consists of a four dimensional hypersurface
embedded in a five-dimensional space which
is {\it asymptotically Minkowski} ( for a recent review of brane models
see \ref{\dick}{R. Dick, \cla18 (2001) R1.}. )
The main motivation for this model is the competition between
the bulk curvature scalar $\cal R$ and the corresponding intrinsic
curvature scalar $R$ on the brane. The action which describes
this particular model is given by
$$
S=M_{5}^{3}\int d^4xdy{\sqrt{-g^{(5)}}}{\cal R}+M_{4}^{2}\int d^4x
{\sqrt{-g^{(4)}}}R
\numbereq\name{\eqtesse}
$$
where $M_5$ stands for the $5-D$ Planck constant and $M_4$ is the
$4-D$ Planck constant. $M_5$ and $M_4$ are independent in general.
The second term of (\eqtesse) is included since matter localized
on the brane might induce it through quantum fluctuations.
This action yields the equations of motion
$$
M_{5}^{3}({\cal R}_{mn}-{1\over 2}{\cal R}g_{mn})+M_{4}^{2}
(R_{\mu\nu}-{1\over 2}Rg_{\mu\nu}){\delta}_{m}^{\mu}
{\delta}_{n}^{\nu}{\delta}(y)
={\delta}_{m}^{\mu}{\delta}_{n}^{\nu}T_{\mu\nu}{\delta}(y).
\numbereq\name{\eqdyo}
$$
In this paper we shall solve the equations (\eqdyo),
with $T_{\mu\nu}=M{\eta}_{\mu\nu}{\delta}^{3}(\vec r)$ in the
linearized approximation and we shall obtain an exact expression
for the gravitational field outside the point source ( linearized
analysis for the Randall-Sundrum model has been performed in
\ref{\lin}{I. Y. Arefe'va, M. Ivanov, W. Muck, K. S. Viswanathan,
and I. V. Volovich, \np590 (2000) 273; Y. S. Myung,
G. Kang and H. W. Lee, \pl478 (2000) 294; H. Collins
and B. Holdom, \pr62 (2000) 124008;
W. Muck, K. S. Viswanathan
and I. V. Volovich, \pr62 (2000) 105019;
R. Dick and D. Mikulovicz, \pl476 (2000) 363.}.)
For ${\vec r} \neq 0$, equation (\eqdyo) implies
$$
{\cal R}=-{2\over 3}{\lambda}R{\delta}(y),
\numbereq\name{\eqtria}
$$
and
$$
{\cal R}_{mn}+{{\lambda}\over 3}g_{mn}R{\delta}(y)+{\lambda}
(R_{\mu\nu}-{1\over 2}Rg_{\mu\nu})
{\delta}_{m}^{\mu}{\delta}_{n}^{\nu}{\delta}(y)=0
\numbereq\name{\eqriddle}
$$
with $\lambda={{M_{4}^{2}}\over {M_{5}^{3}}}$, being a
characteristic length scale of the model.
Here and
throughout the 
paper, we adopt the convention that the Greek indices take values 
0-3 and the Latin indices 0-4. The interesting feature of this scenario is
that although the spectrum of linearized fluctuations does not include
a massless graviton bound state which would correspond to
a $4-D$ graviton it reproduces the correct Newtonian interaction
on the brane for a certain range of parameters
\ref{\dva}{G. Dvali, G. Gabadadze and M. Porrati, \pl484
(2000) 112.}.

In order to describe the real world, the Dvali-Gabadadze-Porrati scenario
has to satisfy all the existing tests of General Relativity.
The most general axially symmetric and static metric in $D=4+1$ 
can always be brought to the following form
$$
ds^2=-e^adt^2+e^bdr^2+e^cr^2d\Omega^2+dy^2,
\numbereq\name{\eqpente}
$$
where $d\Omega^2=d\theta^2+\sin^2\theta d\phi^2$ is the solid angle on 
$S^2$ and $a$, $b$ and $c$ are functions of $r$ and $y$.
The non-zero components of the $4-D$ Ricci tensor for the general
metric (\eqpente) in the linearized approximation are
$$
\eqalign{
R_{tt}&=-{1\over 2}a^{\prime\prime}
-{1\over r}a^\prime\cr
R_{rr}&={1\over 2}a^{\prime\prime}+c^{\prime\prime}
-{1\over r}b^\prime+{2\over r}c^\prime \cr
R_{\theta\theta}&=r^2\Big[{1\over 2}
c^{\prime\prime}+{2\over r}c^\prime+{{a^\prime-b^\prime}\over 2r}
\Big]\cr
R_{\phi\phi}&=R_{\theta\theta}{\sin^2{\theta}}\cr}
\numbereq\name{\eqaigalew}
$$
while the components of the $5-D$ Ricci tensor read
$$
\eqalign{
{\cal R}_{tt}&=R_{tt}
-{1\over 2}\ddot a \cr
{\cal R}_{rr}&=R_{rr}+{1\over 2}\ddot b \cr
{\cal R}_{\theta\theta}&=R_{\theta\theta}
+{{r^2}\over 2}\ddot c\cr
{\cal R}_{\phi\phi}&={\cal R}_{\theta\theta}{\sin^2{\theta}}\cr
{\cal R}_{yy}&={1\over 2}(
\ddot a+\ddot b+2\ddot c) \cr
{\cal R}_{ry}&={\cal R}_{yr}={1\over 2}\Big[
\dot a^\prime+2\dot c^\prime-{2\over r}(\dot b-\dot c)\Big]\cr}
\numbereq\name{\eqakrathtos}
$$
where $a^\prime$ and $\dot a$ indicate differentiation
with respect to $r$ and $y$ respectively.
Consequently the $4-D$ and the $5-D$ Ricci scalars read
$$
\eqalign{
R&=a^{\prime\prime}+2c^{\prime\prime}+{2\over r}a^{\prime}
-{2\over r}b^{\prime}+{6\over r}c^{\prime}+{2\over {r^2}}(b+c) \cr
{\cal R}&=R+{\ddot a}+{\ddot b}
+2{\ddot c}.\cr}
\numbereq\name{\eqdistance}
$$
These equations apply to the positive side of the brane, i.e.,
$y > 0$, the corresponding equations to the negative
side of the brane, $y<0$, are obtained by $Z_2$ symmetry.
The Einstein equations (\eqtria) and (\eqriddle) then demand that
all of the components of ${\cal R}_{mn}$ vanish for $y \neq 0$
and dictate the following matching conditions at $y=0$
$$
\eqalign{
\dot a(r, y)|_{y=0}&=-{{\lambda}\over 2}(a^{\prime\prime}+
{2\over r}a^{\prime})|_{y=0}+{{\lambda}\over 6}R|_{y=0}\cr
\dot b(r, y)|_{y=0}&=-{{\lambda}\over 2}(a^{\prime\prime}+2c^{\prime\prime}
-{2\over r}b^{\prime}+{4\over r}c^{\prime})|_{y=0}
+{{\lambda}\over 6}R|_{y=0} \cr
\dot c(r, y)|_{y=0}&=-{{{\lambda}r^2}\over 2}(c^{\prime\prime}+
{4\over r}b^{\prime}+{{a^{\prime}-b^{\prime}}\over r})|_{y=0}
+{{\lambda}\over 6}R|_{y=0}. \cr}
\numbereq\name{\eqvione}
$$
The $yy$ component of the equation (\eqriddle), together with
equation (\eqtria) implies that $R=0$ for $y \neq 0$, which
leads to $R=0$ at $y=0$ by continuity. This simplifies the
matching conditions (\eqvione)
$$
\eqalign{
\dot a(r, y)|_{y=0}&=-{{\lambda}\over 2}(a^{\prime\prime}+
{2\over r}a^{\prime})|_{y=0}={{\lambda}\over 2}{\ddot a}|_{y=0}\cr
\dot b(r, y)|_{y=0}&=-{{\lambda}\over 2}(a^{\prime\prime}+2c^{\prime\prime}
-{2\over r}b^{\prime}+{4\over r}c^{\prime})|_{y=0}
={{\lambda}\over 2}{\ddot b}|_{y=0} \cr
\dot c(r, y)|_{y=0}&=-{{{\lambda}r^2}\over 2}(c^{\prime\prime}+
{4\over r}b^{\prime}+{{a^{\prime}-b^{\prime}}\over r})|_{y=0}
={{\lambda}\over 2}{\ddot c}|_{y=0}. \cr}
\numbereq\name{\eqvioned}
$$
The solution to the equation ${\cal R}_{tt}=0$ reads
$$
a(r, y)={1\over {2{\pi}^2}}\int_0^\infty dpp^2C(p)j_{0}(pr)
e^{-py}, \qquad y > 0,
\numbereq\name{\eqkoulh}
$$
where $C(p)$ is a function of $p$ to be determined
and $j_0(x)$ is the spherical Bessel 
function.
Substituting (\eqkoulh) into the first of
equations (\eqvione) and taking into account
that $\delta^{3}(\vec r)=\int{{d^3{\vec p}}\over {(2\pi)^3}}e^{i{\vec p}
{\vec r}}={1\over {2{\pi}^2}}\int_0^\infty dpp^2j_{0}(pr)$
we find
$$
C(p)={{C_0}\over {p+{1\over 2}{\lambda}p^2}}.
\numbereq\name{\eqvion}
$$
As a result $a(r, y)$ is given by the expression
$$
a(r, y)={{C_0}\over {2{\pi}^2}}\int_0^\infty dp{{p^2}
\over {p+{1\over 2}{\lambda}p^2}}j_{0}(pr)
e^{-py}, \qquad y>0,
\numbereq\name{\eqkoulhs}
$$
where $C_0$ is a constant to be determined by the requirement
that the solution reproduces the correct Newtonian limit.
The Newtonian limit is specified by the asymptotic
behavior of $g_{00}$ on the brane for large $r$.
Since $p \sim {1\over r}$ and $p+{{\lambda}\over 2}p^2 \sim
{1\over r}+{{\lambda}\over {2r^2}}$ we find that for $r \ll {\lambda}$
$$
a(r, y=0)\simeq {{C_0}\over {{\pi}^2{\lambda}}}\int_0^\infty dp
j_{0}(pr)={{C_0}\over {2{\pi}{\lambda}r}}=-{{2GM}\over r}
\numbereq\name{\eqnewton}
$$
which determines the relation $C_0=-4{\pi}GM{\lambda}$.

The $yy$-component of the equations of motion outside the brane
provide the equation $\ddot a+\ddot b+2\ddot c=0$.
If we impose the following
conditions to the solution
$$
\lim_{|y|, r\to\infty}a=\lim_{|y|, r\to\infty}b
=\lim_{|y|, r\to\infty}c=0
\numbereq\name{\eqexy}
$$
independent of the order of the limits, we find that
$$
a+b+2c=0,
\numbereq\name{\eqvios}
$$
which implies $R=0$ on the brane.
By combining equations ${\cal R}_{ry}=0$ and (\eqvios)
we derive the following expression for $b(r, y)$
$$
b(r, y)=-{1\over {r^3}}{\int^r}dr^{\prime}r^{\prime 2}a(r^\prime, y). 
\numbereq\name{\eqwiotex}
$$
Consquently by substituting (\eqkoulhs) into (\eqwiotex) we find
$$
b(r, y)={{2GM{\lambda}}\over {\pi}}\int_0^\infty dp{{p^2}
\over {p+{1\over 2}{\lambda}p^2}}{{j_{1}(pr)}\over {pr}}
e^{-py}, \qquad y>0.
\numbereq\name{\eqkoulhsh}
$$
Finally equation (\eqvios) provides an expression for $c(r, y)$
$$
c(r, y)={{GM{\lambda}}\over {{\pi}}}\int_0^\infty dp{{p^2}
\over {p+{1\over 2}{\lambda}p^2}}[j_{0}(pr)-{{j_{1}(pr)}\over {pr}}]
e^{-py}, \qquad y>0.
\numbereq\name{\eqkou}
$$
Although $a(r, y)$ satisfies the matching condition,
$b(r, y)$ and $c(r, y)$ fail to do so. The boundary condition
for $b(r, y)$ is (\eqvioned)
$$
{\dot b}(r, y)|_{y=0}=-{\lambda}({1\over 2}a^{\prime\prime}
+c^{\prime\prime}-{1\over r}b^{\prime}+{2\over r}c^{\prime})|_{y=0}
={{\lambda}\over 2}{\ddot b}(r, y)|_{y=0}.
\numbereq\name{\eqasis}
$$
By substituting equation (\eqkoulhsh) into (\eqasis) we find
$$
{{\lambda}\over 2}{\ddot b}(r, y)|_{y=0}-{\dot b}(r, y)|_{y=0}=
{{2GM{\lambda}}\over {{\pi}}}\int_0^\infty dp{p^2}{{j_{1}(pr)}\over {pr}}
={{GM{\lambda}}\over {r^3}}
\numbereq\name{\eqacous}
$$
which is nonzero. The situation is reminiscent with the Randall-Sundrum
model where two sets of solutions to the Einstein equations in two
different coordinate systems, or equivalently, in two different gauges
were constructed. The solution that we found is expressed in coordinates
which are straight with respect to the horizon and is free from
metric singularities far away from the source but it causes the
brane to bend away from $y=0$ \ref{\gar}{J. Garriga and
T. Tanaka, \prl84 (2000) 2778.}. By transforming
the solution to coordinates based on the brane
we shall obtain another solution which satisfies
the matching condition. In this set of coordinates
the brane appears straight. We shall denote the solution in
this gauge by $a^{P}(r, y)$, $b^{P}(r, y)$ and $c^{P}(r, y)$, where
the superscript $P$ means {\it physical}. 

We shall now discuss the gauge symmetries of the Einstein equations,
more specifically the coordinate transformations that respect the
axially symmetric, static form of the five dimensional metric (\eqpente).
Let's perform a coordinate transformation (gauge transformation) generated
by $v, u$, functions of $r$ and $y$ such that
$r \mapsto r+v(r, y)$ and $y \mapsto y+u(r, y)$ and demand that this
particular coordinate transformation respects the form of the
five dimensional metric (\eqpente). We find that the functions
$v, u$ obey to linear order the following relations
$$
{\dot v}=0, \qquad v^{\prime}+{\dot u}=0,
\numbereq\name{\eqrojv}
$$
while the components of the metric transform under these
residual coordinate transformations as follows
$$
{\delta}a=0, \qquad {\delta}b=2u^{\prime},
\qquad {\delta}c=2{{u}\over r}.
\numbereq\name{\eqrpom}
$$
We can easily verify that the Einstein equations remain
invariant under the transformations (\eqrojv).
Thus, the physical solution will be given by
$$
a^{P}=a+{\delta}a, \qquad b^{P}=b+{\delta}b,
\qquad c^{P}=c+{\delta}c.
\numbereq\name{\eqrvmc}
$$
We impose the boundary condition on $b^P$
on the brane,
$$
{{\lambda}\over 2}{\ddot b}^{P}(r, y)|_{y=0}
-{\dot b}^{P}(r, y)|_{y=0}=0.
\numbereq\name{\eqmasoud}
$$
Under a coordinate transformation generated by $u$, $v$ we
find that
$$
{\lambda}{\ddot u}^{\prime}(r, 0)-2{\dot u}^{\prime}(r, 0)=
-{{GM{\lambda}}\over {r^3}}.
\numbereq\name{\eqreion}
$$
The parameters of the transformation, $u$ and $v$, satisfy equations
(\eqrojv). Consequently,
$$
v=v(r), \qquad u(r, y)=-v^{\prime}(r)y+{\chi}(r),
\numbereq\name{\eqweis}
$$
where $\chi(r)$ is an arbitrary function of $r$.
By substituting (\eqweis) into (\eqreion) we find that
$$
v(r)=-{{GM{\lambda}}\over {4r}}, \qquad u(r, y)
=-{{GM{\lambda}}\over {4r^2}}y
+{\chi}(r).
\numbereq\name{\eqvret}
$$
The physical solution to the first order in the gravitational
coupling takes then the form
$$
\eqalign{
a^{P}(r, y)&=-{{2GM{\lambda}}\over {{\pi}}}\int_0^\infty dp{{p^2}
\over {p+{1\over 2}{\lambda}p^2}}j_{0}(pr)
e^{-py}\cr
b^{P}(r, y)&={{2GM{\lambda}}\over {{\pi}}}\int_0^\infty dp{{p^2}
\over {p+{1\over 2}{\lambda}p^2}}{{j_{1}(pr)}\over {pr}}
e^{-py}+{{GM{\lambda}}\over {r^3}}y+2{\chi}^{\prime}(r)\cr
c^{P}(r, y)&={{GM{\lambda}}\over {{\pi}}}\int_0^\infty dp{{p^2}
\over {p+{1\over 2}{\lambda}p^2}}[j_{0}(pr)-{{j_{1}(pr)}\over {pr}}]
e^{-py}-{{GM{\lambda}}\over {2r^3}}y
+{{2{\chi}(r)}\over r}. \cr}
\numbereq\name{\eqerions}
$$
Finally for $2GM \ll r, y \ll {\lambda}$ the solution takes the form
$$
\eqalign{
a(r, y)&=-{{2GM}\over {i{\pi}r}}{\ln{{y+ir}\over {y-ir}}}=
-{{4GM}\over {{\pi}r}}{\tan^{-1}{r\over y}}=-{{2GM}\over r}
(1-{2\over {\pi}}{\tan^{-1}{r\over y}})\cr
b(r, y)& \cong {{GM}\over r}\Big [ (1-{{y^2}\over {r^2}})(1
-{2\over {\pi}}{\tan^{-1}{r\over y}})+{{2y}\over {{\pi}r}}\Big ]\cr
c(r, y)&=-{1\over 2}(a+b) \cong {{GM}\over 2r}\Big [ (1+{{y^2}\over {r^2}})
(1-{2\over {\pi}}{\tan^{-1}{r\over y}})-{{2y}\over {{\pi}r}}\Big ]\cr}
\numbereq\name{\eqjoey}
$$
We observe that if we confine the metric on the
brane $y=0$, the {\it physical solution} reads for
$2GM \ll r \ll {\lambda}$
$$
a^{P}(r)=-{{2GM}\over r}, \qquad b^{P}(r)={{GM}\over r}+2{\chi}^{\prime},
\qquad c^{P}(r)={{GM}\over {2r}}+{{2\chi}\over r},
\numbereq\name{\eqvior}
$$
where $\chi$ is an arbitrary function of $r$. In order to
recover the standard static, spherically symmetric
metric on the brane, i.e. $c^{P}=0$ at $y=0$ we choose
$\chi=-{{GM}\over 4}$. This leaves
$b^{P}={{GM}\over r}$, which differs from the standard form of the
Schwarzschild metric and thus leads to different predictions
for the bending of light. In the Randall-Sundrum model the choice
of $\chi$ that sets $c^{P}=0$ leads also to $b^{P}={{GM}\over 2r}$
which is the form of the Schwarzschild metric in the linearized
approximation.

Similarly for $r, y \gg {\lambda}$
$$
\eqalign{
a(r, y)& \cong -{{2GM{\lambda}}\over {r^2+y^2}} \cr
b(r, y)& \cong {{2GM{\lambda}}\over {r^2}}(1-{y\over r}
{\tan^{-1}{r\over y}}) \cr
c(r, y)&=-{1\over 2}(a+b) \cong
{{GM{\lambda}}\over {{\pi}r^2}}({{y^2}\over {y^2+r^2}}-{y\over r}
{\tan^{-1}{r\over y}}). \cr}
\numbereq\name{\eqarvitas}
$$
We note that if we confine the metric on the {\it physical brane}
$y=0$, the solution reads
$$
a^{P}(r)=-{{2GM{\lambda}}\over {r^2}}, \qquad b^{P}(r)
={{2GM{\lambda}}\over {r^2}},
\qquad c^{P}(r)=0,
\numbereq\name{\eqviorio}
$$
which agrees with the standard linearized form of the
$5-D$ Schwarzschild metric.

In this paragraph we shall recapitulate what we have done in this
paper. We solved the linearized
Einstein equations in the model proposed by
Dvali, Gabadadze and Porrati with a static, spherically symmetric
matter distribution on the {\it physical brane} and found an exact
expression for the gravitational field. Although when confined on
the {\it physical brane} the metric reproduces the $4-D$
Newtonian potential for a certain range of parameters,
more specifically for $2GM \ll r \ll {\lambda}$
it fails to reproduce the standard Schwarzschild
metric in its linearized form. Does this result then imply that the
model is incompatible with the experimental predictions of
general relativity? Recently in \ref{\def}{C. Deffayet, G. Dvali,
G. Gabadadze and A. Vainshtein, {\it " Nonperturbative Continuity in
Graviton Mass versus Perturbative Discontinuity "}, hep-th/0106001.}
it was argued that the linearized approximation ceases to be reliable
for this range of parameters and subsequently
an exact classical solution to all orders in the gravitational
coupling is needed. This argument is based on an early work by Vainshtein
\ref{\van}{A. Vainshtein, \pl39 (1972) 393.} in which he suggested
that the limit of vanishing mass of massive general relativity
may coincide with the predictions of massless general relativity
in contrast with previous papers \ref{\van}{H. van Dam and M. Veltman,
\np22 (1970) 397}, \ref{\zak}{V. I. Zakharov, \jetp12 (1970) 312.}
if the nonperturbative exact classical solution is used.
The analog of the graviton mass on the brane is the inverse of the
length scale ${\lambda}$. The possibility then remains that a nonperturbative
solution exists and it reproduces the standard Schwarzschild solution
when confined on the {\it physical brane}.
On the contrary the metric
for $r \gg \lambda$ behaves as a $5-D$ Schwarzschild metric.
(See  also \ref{\geor}{J. Avery, R. Mahurin and G. Siopsis,
{\it " A Brane in five-dimensional infinite flat Minkowski space "},
hep-th/0108132.} where it is claimed that four-dimensional behavior
is recovered at large distances in the Dvali-Gabadadze-Porrati model.)
We hope to report in the near future results which deal with these issues.

Beyond the linear approximation, the properties of an event
horizon which we established in a rigorous manner
for the Randall-Sundrum model
\ref{\egw}{I. Giannakis and H. c. Ren, \pr63 (2001) 125017.}
apply to the DGP brane model as well. If a Schwarzschild
black hole localized on the brane exists, the extension of the
horizon into the bulk with the Gauss normal form of the metric (\eqpente)
is tubular. This is not in contradiction with the analysis
based on the linearized
approximation, employed in this work, since the approximation breaks down
for ${\lambda}y \gg r^2$, thus permiting the horizon to extend to
$y \mapsto \infty$.

\vskip .1in
\noindent
{\bf Acknowledgments.} \vskip .01in \noindent

We would like also to thank A. Polychronakos, M. Porrati
and G. Siopsis for useful discussions.
This work was supported in part by the Department of Energy under Contract
Number DE-FG02-91ER40651-TASK B.

\immediate\closeout1
\bigbreak\bigskip
\line{\twelvebf References. \hfil}
\nobreak\medskip\vskip\parskip

\input refs

\vfil\end

\bye